\documentclass[aps,12pt,preprintnumbers,nofootinbib]{revtex4}

\usepackage{graphicx}
\usepackage{amsmath,amssymb,epsfig,bm,comment}
\usepackage{fancyhdr}
\usepackage[dvips]{color}
\usepackage{mathrsfs}

\definecolor{gray}{rgb}{0.4,0.4,0.4}

\def\({\left(}
\def\){\right)}
\def\[{\left[}
\def\]{\right]}
\def\<{\langle}
\def\>{\rangle}

\newcommand\half{{\ensuremath{\frac{1}{2}}}}
\newcommand\p{\ensuremath{\partial}}

\newcommand{\be}{\begin{equation}}
\newcommand{\ee}{\end{equation}}
\newcommand{\bea}{\begin{eqnarray}}
\newcommand{\eea}{\end{eqnarray}}
\newcommand{\bwt}{\begin{widetext}}
\newcommand{\ewt}{\end{widetext}}
\newcommand{\nn}{\nonumber\\}
\newcommand{\bi}{\begin{itemize}}
\newcommand{\ei}{\end{itemize}}
\newcommand{\ben}{\begin{enumerate}}
\newcommand{\een}{\end{enumerate}}
\newcommand{\bca}{\begin{cases}}
\newcommand{\eca}{\end{cases}}
\newcommand{\bln}{\begin{align}}
\newcommand{\eln}{\end{align}}
\newcommand{\bst}{\begin{split}}
\newcommand{\est}{\end{split}}

\newcommand\ep{\epsilon}

\newcommand\ov{\over}

\begin{document}
\title{Mixed RG Flows and Hydrodynamics at Finite \\Holographic Screen}

\author{Yoshinori~Matsuo$^{\ddagger}$, Sang-Jin~Sin$^{\dagger}$ and Yang~Zhou$^{*}$}
\affiliation{
$\ddagger$ High Energy Accelerator Research Organization(KEK),
Tsukuba, Ibaraki 305-0801, Japan\\
$\dagger$ Department of Physics, Hanyang University,
Seoul 133-791, Korea\\
$*$ Center for Quantum
Spacetime, Sogang University, Seoul 121-742, Korea}
\begin{abstract}
We consider quark-gluon plasma with  chemical potential and
study  renormalization group flows of transport coefficients in the framework of gauge/gravity duality. 
We first study them using the flow equations
and compare the results with hydrodynamic results by  calculating the Green functions on the arbitrary slice.
Two results match exactly.  Transport coefficients at arbitrary scale is ontained by calculating hydrodynamics Green functions.  
When either  momentum or charge vanishes, transport coefficients decouple from each other.  
\end{abstract}
\maketitle
\section{Introduction}

In the application of  AdS/CFT \cite{Maldacena:1997re},   calculations are usually done on the holographic screen at infinity.
However, according to the renormalization group (RG) ideas~\cite{kraus, Susskind, verlinde}, those
are the UV fixed point values which can not be reached by experiments 
performed at the finite energy scale.
Therefore we need to run them down to the scale where one actually performs the experiment by the renormalization group  flow.
 Recent studies of Wilsonian approach~\cite{Joe.RG, Hong.RG} to holographic RG flow of transport
coefficients~\cite{Sin:2011yh} in the framework of gauge/gravity duality suggest that  those apparently different
 approaches of sliding membrane~\cite{Hong.Membrane}, Wilsonian fluid/gravity~\cite{Andy.RG} and
holographic Wilsonian renormalization group~(HWRG)\cite{Joe.RG,Hong.RG,Wilson:RG} are equivalent~\cite{Sin:2011yh}.
Some of the transport coefficients such
as shear viscosity $\eta$, AC conductivity  and diffusion constant $D$ have non-trivial radial flows so that they interpolate the horizon values~\cite{son2} and the boundary values~\cite{son1} smoothly. The holographic Wilsonian RG is also useful to understand the low energy effective theory for holographic liquid~\cite{Faulkner:2009wj,Joe.Semiholography,Son.Deconstruction}. 

Since the discussions so far has been mostly for zero charge cases, one natural question is how to extend it to the case with
 finite  chemical potentials. Unlike the zero charge case, metric fluctuation and Maxwell
fluctuation will mix in charged black hole background.
In general, the mixing effect of metric and Maxwell fluctuations in charged black hole is of essential importance, since it is the reason why
transverse vector modes of Maxwell fields can diffuse and
longitudinal Maxwell modes can have
sound modes  in the presence of the charge.
Part of the answer has  been given in~\cite{Andy.RG}  where
the cutoff dependence of diffusion constant for the shear part of metric
perturbations was calculated. The authors achieved it without
explicit decoupling procedure  by taking
a specific scaling  which,  somehow, effectively  decouples the mixed  modes.
However, it is not clear how to understand why it happens and what scaling limit should be taken for other (i.g. sound) modes.

 In this paper, we first establish the
 flow equations of transport coefficients in the presence of chemical potential   and  numerically integrate the flow equations.
We then calculate transport coefficients directly  by hydrodynamic calculation on the  holographic screen at  finite radial position and compare the two results.
We will find complete agreement.

This paper is organized as follows. In section \ref{equivalence}, we will briefly review the holographic RG flow of transport coefficients.  We also derive the running diffusion constant in zero charged black hole from sliding membrane which reproduces the previous
result in~\cite{Andy.RG} obtained by scaling method.
In section \ref{chargedBH}, we set up the charged black hole medium and focus on mixing structure of metric and Maxwell perturbations. We emphasize on organizing the coupled equations of motion. In section \ref{mixedflow}, we first write down the decoupled flow equation for both electric conductivity and ``conductivity" for momentum current at zero momentum. And then we worked out the ``conductivity" for momentum current in diffusion scaling limit following~\cite{Andy.RG}. We write down the complete mixed flow equations for electric conductivity, momentum current ``conductivity'' and mixing parameter in the end of this section.  In section \ref{hydro}, we use the hydrodynamical method to calculate the Green functions at finite screen $r=r_c$.  We obtained the transport coefficients from Green functions by Kubo formula and we found complete agreement with  the  results from flow equations in section \ref{mixedflow}.  We conclude in last section. 


\section{ Holographic Renormalization Group and Running transport coefficients: A review}\label{equivalence}
In this section, we shall discuss  a few  approaches  to RG flow of transport coefficients and discuss the equivalence between them: They are
 sliding membrane paradigm~\cite{Hong.Membrane}, Wilsonian fluid/gravity~\cite{Andy.RG} and
Holographic Wilsonian RG~\cite{Joe.RG,Hong.RG}.
The radial flow  for transport coefficients can  be derived from holographic Wilsonian RG equation. It can also be derived from the classical equation of motion.  The idea of membrane paradigm is to consider a
membrane action~\cite{Parikh:1997ma} coming from the boundary action at sketched horizon. Similarly  sliding membrane~\cite{Hong.Membrane} paradigm   uses the boundary action at arbitrary slice. From that action, one can   obtain the retarded Green function by solving the equation of motion
perturbatively in hydrodynamic regime. The linear response theory then gives the transport coefficients. The Green functions and  transport coefficients satisfy the same radial evolution  coming from holographic Wilsonian effective action $S_B$, which
in turn  can be obtained from integrating out UV geometry directly~\cite{Hong.RG,Sin:2011yh}. $S_B$ can also be treated as a
boundary action at cutoff slice and it induces multi-trace deformations for IR dynamics~\cite{Joe.RG,Hong.RG}. For more discussions about the essential multi-trace deformations see~\cite{Witten:2001ua,Vecchi:2010dd}. If one define the linear level
transport coefficient at cut off slice, then the Hamilton-Jacobi equation for the effective action can directly give radial
flow of them. These flows are exactly the same as those coming from equations
of motion as explicitly shown in~\cite{Sin:2011yh}.


Technically, sliding membrane paradigm is more convenient than holographic Wilsonian RG simply because the former
starts from equation of motion while the latter is conceptually more
satisfactory.    We shall mostly use
the former in this paper.

We briefly review the sliding membrane paradigm~\cite{Hong.Membrane} below. 
We derive the cutoff dependent diffusion constant, which is also calculated from the Wilsonian approach of Fluid/Gravity~\cite{Andy.RG}. 
We show that the result from the sliding membrane paradigm reproduces 
that in \cite{Andy.RG} to demonstrate the equivalence of sliding membrane paradigm~\cite{Hong.Membrane}   and Wilsonian approach of
Fluid/Gravity~\cite{Andy.RG}. 
We start with standard Maxwell action \be S=-\int
d^{d+1}x\sqrt{-g}{1\ov 4g_{\text{eff}}^2(r)}F_{MN}F^{MN}\ , \ee with the background metric \be ds^2 = -g_{tt}dt^2 +
g_{rr}dr^2 + g_{ii}dx^idx^i\ . \ee  Defining 
$J^\mu$  and $G$   by  \be J^\mu = -{1\ov g_{\text{eff}}^2}\sqrt{-g}F^{r\mu}\
,\quad G = \sqrt{-g}/g_{\text{eff}}^2\ , \ee
equations of motion can be written as~\footnote{Here we follow the notation in~\cite{Hong.Membrane}.} 
\bea
\p_tJ^t + \p_z J^z &=& 0 \label{conservation}\ ,\\
\p_rJ^t + Gg^{tt}g^{zz}\p_zF_{zt} &=& 0\ ,\\
\p_rJ^z - Gg^{tt}g^{zz}\p_tF_{zt} &=& 0\label{zEom1}\ , \eea where $z$ is the momentum direction  and we focus on the
longitudinal mode first. 
The  Bianchi identities are given by:
\be
-{g_{rr}g_{zz}\ov G}\p_tJ^z - {g_{rr}g_{tt}\ov G}\p_zJ^t + \p_rF_{zt} = 0\ .\label{Bianchi} \ee
Using the definition of the conductivity $\sigma=J^z/E_z=J^z/F_{zt}$,
we have   \be \p_r\sigma = {\p_rJ^z F_{zt} - J^z
\p_r(F_{zt})\ov F_{zt}^2}\ ,\quad \sigma^2 = {(J^z)^2\ov(F_{zt})^2}\ .\label{sigma1} \ee We want to replace all the
$\partial_r$ terms  in above equation   using equations of motion.
 Taking use of ($\ref{conservation}$),
($\ref{zEom1}$) and ($\ref{Bianchi}$), all the currents and fields disappear and finally we have the flow for
conductivity~\cite{Hong.Membrane} \be\label{sigmaflow0} {\p_r\sigma\ov i\omega} = g^2_{\text{eff}}\sigma^2\left[{1\ov
\sqrt{-g}g^{rr}g^{zz}} - {k^2\ov \omega^2}{1\ov \sqrt{-g}g^{rr}g^{tt}}\right] - {1\ov g^2_{\text{eff}}}
\sqrt{-g}g^{tt}g^{zz}\ . \ee Before doing explicit analysis for this conductivity flow we should mention that there are
two important scaling limits for the frequency dependent flow equations: one is $\omega\sim k^2<<1$ and the other is
$k= 0, \omega<<1$. In the diffusion scaling regime $\omega\sim k^2<<1$, we have following solution of
($\ref{sigmaflow0}$) for conductivity : \be\label{flowsigma} {1\ov \sigma(r_c)} = {1\ov \sigma_H} -{k^2\ov
i\omega}{1\ov f_0}\ ,\quad\quad {1\ov f_0} = \int_{r_ H}^{r_c}{g^2_{\text{eff}}\ov \sqrt{-g}g^{rr}g^{tt}}\ . \ee Since
the (complex)  conductivity
 at arbitrary slice and the green function is related by  \be \sigma^{ij}(k_\mu, r_c) = -{G_R^{ij}({k_\mu, r_c})\ov i\omega}\ , \ee we can write  the Green
function at   cutoff surface in terms of horizon conductivity  $\sigma_H$
\be G_R^{ii}({k_\mu, r_c}) = {\omega^2\sigma_H\ov
i\omega-D(r_c) k^2}\ , \ee where
  $D(r_c)=\sigma_H/f_0\ $
  is the  cutoff dependent diffusion constant.
  Substituting
into ($\ref{flowsigma}$), we have\footnote{Remember that in this formula, the conductivity is defined using \be
\sigma(z_c) ={J^i\ov E_i} = {\sqrt{-g}F^{ri}\ov -F_{ti}}\ . \ee For a physical observer hovering at $z_c$ surface we
want to use the following normalizations \be \sigma(z_c)\rightarrow \hat\sigma(z_c)\equiv{\sigma(z_c)\ov
(g_{ii})^{d-3\ov 2}}\ . \ee} 
 \be {1\ov \sigma(r_c)} = {1\ov \sigma_H} -{k^2\ov i\omega}{D(r_c)\ov \sigma_H}\ . \ee
In the orthonormal frame, we have the normalized momentum as follows\be \omega\rightarrow \omega_c \equiv {\omega\ov
\sqrt{g_{tt}}}\ ,\quad k\rightarrow k_c\equiv{k\ov \sqrt{g_{ii}}}\ . \ee Using these new variables we can rewrite
($\ref{flowsigma}$) as \be {1\ov \sigma(r_c)} = {1\ov \sigma_H} -{k_c^2\ov i\omega_c}{\hat D(r_c)\ov \sigma_H}\ , \ee
where $\hat D(r_c)$ is given by \be \hat D(r_c)\equiv D(r_c){g_{ii}\ov \sqrt{g_{tt}}}\ . \ee

Defining the local temperature  by  $ T_c={T_H\ov \sqrt{g_{tt}}}\ ,$
and a dimensionless diffusion constant  by
   $\bar D_c\equiv \hat DT_c$,
  we have \be {\bar D_c\ov T_c} = {\sigma_H g_{ii}\ov f_0 \sqrt{g_{tt}}}\ . \ee This is precisely the result obtained in~\cite{Andy.RG},
  supporting the equivalence of two approaches.
So far we considered only chargeless case. It is known that 
presence of charge introduces non-trivial mixing between modes. 
We will consider what happens for the flow of {\it charged} black hole below.

\section{Mode mixing in Charged AdS Black hole}\label{chargedBH}
In order to describe low energy physics of various strongly coupled systems, we need different IR deformations for the
$d+1$ dimensional AdS background.
Consider a $d$ dimensional holographic system with finite charge density. The minimal $d+1$ bulk action is
\be\label{RNaction} S = \int d^{d+1}x\sqrt{-g}\left[{1\ov 2\kappa^2}\left( {R} + {d(d-1)\ov L^2}\right) - {1\ov
4g^2}F^2\right]\ , \ee where Newton constant $G_N=\kappa^2/8\pi$ and $g$ is the Maxwell coupling. Maxwell and   Einstein equations are given by
\bea\label{EOM}
 \p_\mu(\sqrt{-g}F^{\mu\nu})&=& 0, \;\; \\
{1\ov
\kappa^{2}}\left(R_{\mu\nu}-{1\ov 2}g_{\mu\nu} {R} -g_{\mu\nu}{d\ (d-1)\ov 2L^2}\right) &=&{1\ov
2g^2}\left(2F_{\mu\rho}F_\nu^\rho-{g_{\mu\nu}\ov 2}F_{\rho\sigma}F^{\rho\sigma}\right)\ . \eea
The charged black hole solution for this action is \bea ds^2 &=& {r^2\ov L^2}\left(-f(r)dt^2 +  dx^idx^i \right) +
{L^2\ov r^2 } {dr^2\ov f(r)}\ ,\label{metric}\\
\bar A_t &=& \mu\left(1-{r_0^{d-2}\ov r^{d-2}}\right)\ ,\quad f(r) = 1 + \alpha Q^2{r_0^{2d-2}\ov r^{2d-2}} - (1+\alpha
Q^2){r_0^d\ov r^d}\ .\label{metric1} \eea where $\alpha={L^2\kappa^2\ov (d-1)(d-2)g^2}$ is a dimensionless coupling.
Chemical potential $\mu$ is related to $Q$ by $Q={\mu(d-2)\ov r_0}$ and the horizon $r_0$ is the largest root of
$f(r)=0$.

\subsection{Modes in RN-AdS} 
We consider  the
perturbations   \be g_{\mu\nu}\rightarrow g_{\mu\nu}+\delta g_{\mu\nu} ,\quad A_\mu \rightarrow
A_\mu+\delta A_\mu\ ,\ee  around the background  $g_{\mu\nu}$ and $A_\mu$, which are given in $(\ref{metric})$ and
$(\ref{metric1})$. One can use background metric raise and lower tensor indices.
 The linearized gravity fluctuations
can be decomposed into tensor, vector and scalar type~\cite{Kodama:2000fa, Kovtun:2005ev}. We will consider the first two types
and leave the last one elsewhere. In the charged black hole background, vector part of  metric perturbation and transverse Maxwell perturbation will couple.
We can organize the the former such that its action and equations
 look like  that of longitudinal Maxwell perturbation. For
more details about the action of linearized fluctuations in RN-AdS, see~\cite{Ge:2008ak,Edalati:2010hk}.

\subsubsection{Tensor Mode $\delta g^x_y$}
 By symmetry argument or by direct analysis
  one can find that  off-diagonal perturbation $\phi\equiv \delta g^x_y$ decouples from all other perturbations. Thus it satisfies the
same equation of motion to that for a minimally coupled massless scalar:
 \be
\p_\mu (\sqrt{-g} \p^\mu\phi)=0\ . \ee
This evolution for $\phi$ will give the flow of correlation of the corresponding
operator $T_{x}^y$. Notice that if we work in $d=3$, we should take $k\rightarrow0$ limit since we only have two spatial directions.

 \subsubsection{Vector Modes $\delta g^x_{z}, \delta g^x_{t}, \delta g^x_r, \delta A_x$}
 Picking up a momentum  parallel to $z$ direction, we can only have $d-2$ transverse dimension left. One can observe that the shear mode of metric perturbations of components $g_{xz}, g_{xt}, g_{xr}$ which behave similarly as Maxwell field $a_z, a_t, a_r$~\cite{son2,Hong.Membrane} defined by
 \be
 a_z\equiv \delta g^x_{z},\quad a_t\equiv \delta g^x_{t}, \quad a_r\equiv \delta g^x_r\ .
 \ee
 In terms of these variables, finally the vector part off shell action in $d+1$ dimensional RN-AdS can be written as follows:\footnote{We derive this action up to total derivative terms, which are irrelevant to equations of motion. See Appendix A for details.}
\be
S=\int d^{d+1}x\; \sqrt{-g} \left( -\frac{1}{4g^2 }  F^{\mu\nu}F_{\mu\nu} 
        - \frac{1}{4g^2_{\text{eff}}(r)}  f^{\mu\nu}f_{\mu\nu}   +\frac{1}{g^2}a_t A'_x{\bar A}'_t \right)\ .
\ee
Here we have used $j$ and $f$ to denote the current and strength for the $a_\mu$:\be
j^\mu = -{1\ov g_{\text{eff}}^2(r)}\sqrt{-g}f^{r\mu}\ ,\quad f_{\mu\nu} = \p_\mu a_\nu - \p_\nu a_\mu\ ,
\ee and $J$ and $F$ to denote the current and strength for $A_\mu$. Consider first the chargeless case, 
 equations of motion for shear part metric fluctuations can be written as
\bea
\p_tj^t + \p_z j^z &=& 0 \\
\p_rj^t + Gg^{tt}g^{zz}\p_zf_{zt} &=& 0\\
\p_rj^z - Gg^{tt}g^{zz}\p_tf_{zt} &=& 0\ ,
\eea
where the effective determinant and the effective coupling are
\be
G = \sqrt{-g}/g^2_{\text{eff}}(r)\ ,\quad {1\ov g^2_{\text{eff}}(r)}= {g_{xx}\ov 2\kappa^2}\ .
\ee
For the charged-AdS black hole, we introduce the new charge density which is defined as
\be
\tilde{j^t} =  j^t - \frac1{g^2}\sqrt{-g}{\bar A}'_t A_x\ . 
\ee
The equations of motion takes the same form as chargeless case, 
\bea
\p_t\tilde{j^t} + \p_z j^z &=& 0\label{heom0} \\
\p_r\tilde{j^t} + Gg^{tt}g^{zz}\p_zf_{zt} &=& 0 \label{heom1}\\
\p_rj^z - Gg^{tt}g^{zz}\p_tf_{zt} &=& 0\label{heom2}\ ,
\eea
but $j^t$ is replaced by $\tilde j^t$, 
while the Bianchi identity holds as before
\be\label{heom3}
-{g_{rr}g_{zz}\ov G}\p_tj^z - {g_{rr}g_{tt}\ov G}\p_zj^t + \p_rf_{zt} = 0\ .
\ee
 Defining $J^x$ and $\tilde J^x$ by 
 \be
J^x = -\frac1{g^2}\sqrt{-g}g^{rr}g^{xx}\p_r A_x\ , \quad 
  \tilde{J}^x = J^x + \frac1{g^2}\sqrt{-g}  {\bar A}'_t  a_t\ee
 the equation of motion for $A_x$ can be written as
\be\label{Aeom} -\p_r \tilde{J^x} + \frac1{g^2}\sqrt{-g}g^{xx}(-g^{tt}\p_t F_{tx} + g^{zz}\p_z F_{zx}) = 0\ . \ee
\subsubsection{Vector Mode $A_x$ at $k_z=0$}
From (\ref{Aeom}) and (\ref{heom1}), one can see that 
  $  A_x$   decouples from $a_t$ in the $k_z\to 0$  limit: 
\be\label{EOMAxD}
\p_r(\sqrt{-g}g^{rr}g^{xx}\p_r  A_x) - \frac1{g^2}\sqrt{-g}g^2_{\text{eff}}({\bar A}'_t)^2 A_x +
\omega^2\sqrt{-g}g^{xx}g^{tt} A_x =0\ . \ee 
The relevant on-shell action for $A_x$ at boundary $r=r_c$ can be written as 
 \be\label{AonshellS} S_{\text{on-shell}} =  \int_{r=r_c} \tilde J^x A_x\ . \ee

 
 \section{Flows of Transport Coefficients}\label{mixedflow}
We shall establish the transport coefficient flows for the charged black hole in this section.
\subsection{ Shear Viscosity Flow and Scalar Response}
Since tensor mode decouples from all other perturbations, it behaves as a massless scalar perturbation even in the charged black hole background. Following the
sliding membrane argument~\cite{Hong.Membrane}, it is useful to define a cutoff dependent scalar response function \be
G_\phi (r_c, k_\mu) ={ -\sqrt{-g}g^{rr}\p_r\phi\ov 2\kappa^2 \phi(r_c, k_\mu)}\ . \ee And we define \be \eta(r_c, k_\mu):=
{G_\phi (r_c, k_\mu)\ov i\omega} \ee which satisfies a flow equation evolving equation of motion at $k=0$: \be\label{etaflow}
{\p_{r_c}\eta(r_c, \omega)}=i\omega \left({2\kappa^2 \eta^2(r_c, \omega)\ov \sqrt{-g}g^{rr}}-\frac{\sqrt{-g}g^{tt}}{2\kappa^2}\right)\ . \ee It is
manifest from above equation that in the zero frequency limit, $\eta(r_c)$ is independent of $r_c$. Its value is request to be  
\be\label{floweta}
\eta(r_c)=\eta(r_0)={1\over 16\pi G}\left(\frac{r_0}{L}\right)^{d-1},\ee
by the horizon regularity. 
Since the entropy density is $s=\frac{1}{4G} \left(\frac{r_0}{L}\right)^{d-1}$, the ratio $\eta/s$  does not run. 
In the next section we will see that this is consistent with the 
direct calculation of $\eta(r_c)$ using the holographic hydrodynamics at the 
 finite holographic screen.%
\footnote{In reference \cite{Andy.RG}, the entropy density is defined to be proportional to  one over  embedded volume of the cutoff membrane and $\eta$ is also 
  defined  that 
$\eta/s$ does not run. As a consequence both their entropy density and  viscosity run.   Also both vanish  at the 
infinity. }
 
\subsection{ Electric conductivity flow  at the zero momentum }\label{sigmaA1}
\begin{figure}[hbtp]
\includegraphics*[bb=0 0 250 160,width=0.45\columnwidth]{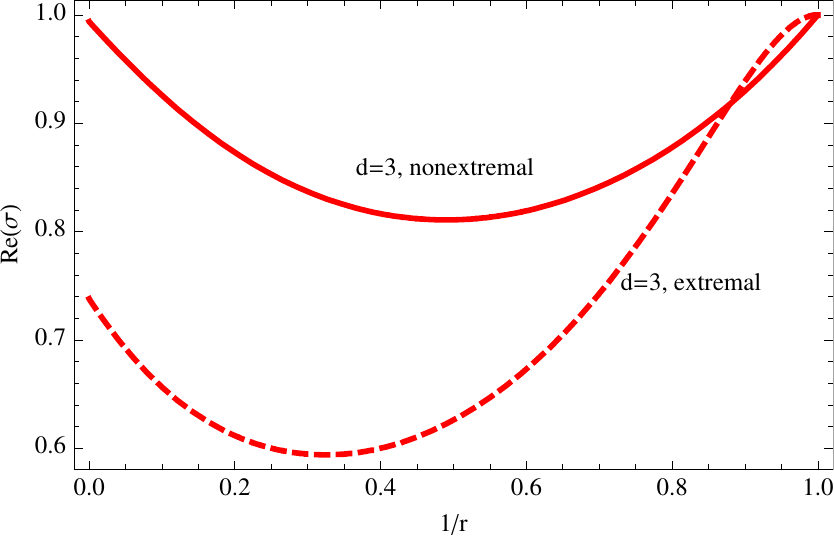}~~~~~~~~
\includegraphics*[bb=0 0 250 160,width=0.45\columnwidth]{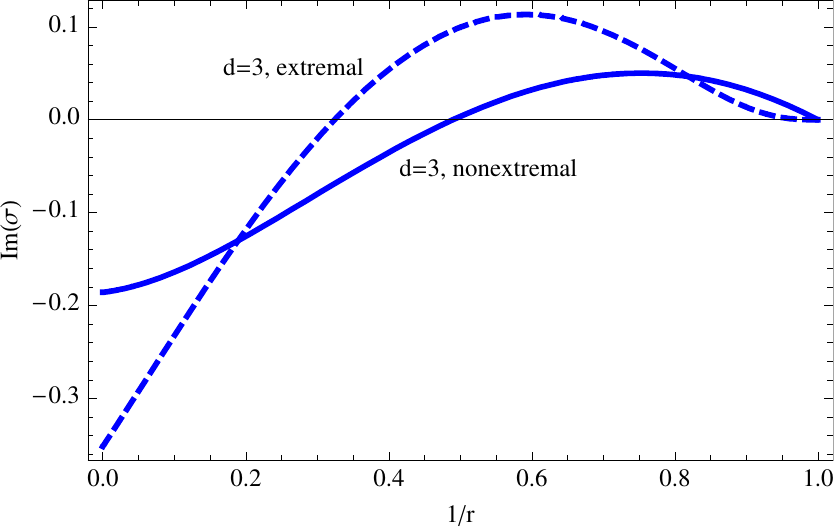}
\caption{ $r$ flow of AC conductivity, with $d=3$  RN-AdS black hole background. Here we normalize $r$ such that the horizon localizes at $r=1$
.
  \label{fig1}}
\end{figure}

 Now we shall consider the conductivity flow coming from $ A_x$ perturbation with zero momentum.
From (\ref{AonshellS}) one can obtain the conjugate momentum for $ A_x(r)$ at $r=r_c$ \be\label{CM} \Pi_{  A_x}(r_c) = {\delta S_{\text{on-shell}}\ov \delta\   A_x } = J^x + \frac1{g^2}\sqrt{-g}{\bar A}_t' a_t \ . \ee
Notice that this is nothing but the shifted current ${\tilde J}^x $ which 
was introduced previously to simplify the equation of motion. 
As explained in~\cite{Hartnoll:2009sz}, the first and second term
in right hand side of (\ref{CM}) should be related to electric conductivity and thermal-electric conductivity
respectively. Notice that the  on-shell action  is the integral over the membrane  at   $r=r_c$ and the conjugate momentum
eq.(\ref{CM}) is also defined at  $r_c>r_0$. 
 In the limit of zero momentum, $A_x$ decouples and  we can define  electric conductivity by 
\be\label{sigmadef} \sigma_A(\omega,r_c) := {J^x\ov i\omega A_x}\ .\ee One can quickly rewrite the equation of motion (\ref{EOMAxD}) as flow equation for $\sigma_A$:
\bea\label{sigmaA0flow}
{\p_{r_c} \sigma_A\ov -i\omega} + { g^2\sigma_A^2\ov \sqrt{-g}g^{rr}g^{xx}} + { g^2_{\text{eff}}\sqrt{-g}(\bar A'_t)^2\ov {g^2}\omega^2}
- \frac1{g^2}\sqrt{-g}g^{xx}g^{tt} = 0\ .
\eea
 Due to the regularity condition at the horizon, we
need \be\label{sigmaH} \sigma_H\equiv\sigma_A(r_0) = {1\ov g^2}\left({r_0\ov L}\right)^{d-3}\ , \ee which is consistent with horizon conductivity evaluated in~\cite{Hong.Membrane}. In order to see the
solutions explicitly, we plot the conductivity flows with different black hole charges in $d=3$ in Figure
\ref{fig1}.

 A few remarks are in order for these flow solutions.
 
\begin{figure}[hbtp]
\includegraphics*[bb=0 0 250 160,width=0.45\columnwidth]{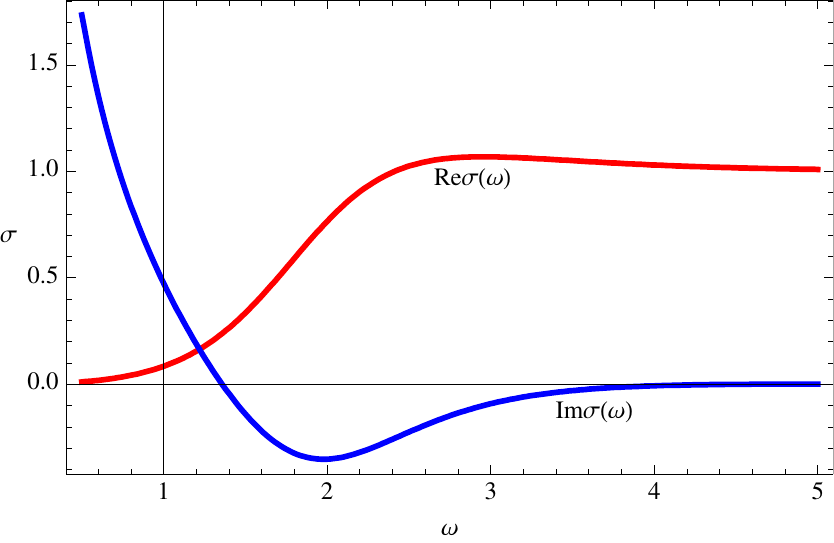}~~~~~~~~~
\includegraphics*[bb=0 0 250 160,width=0.45\columnwidth]{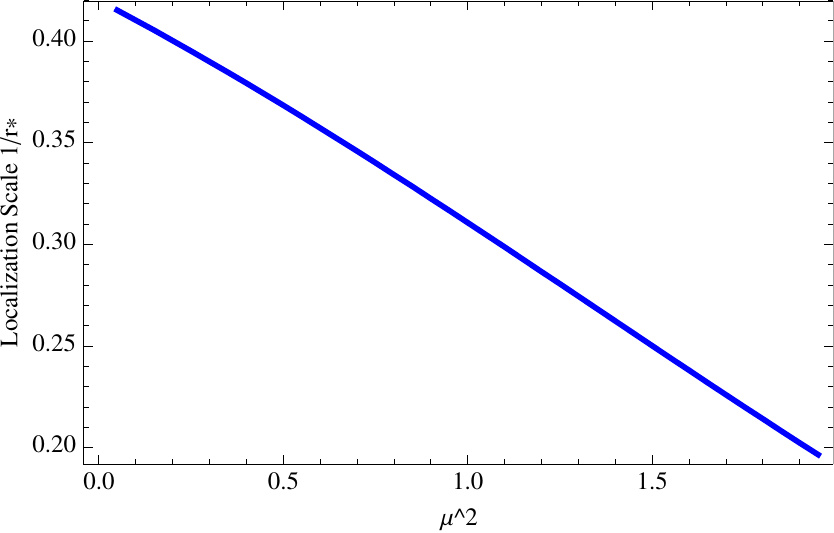}
\caption{Left: Frequency dependence of boundary AC conductivity, with $d=3$ RN-AdS black hole background .
As a  check for our flow solutions, the boundary AC conductivity is precisely consistent with the results in ~\cite{Hartnoll:2009sz}. Right: Chemical potential dependence of scale $r_*$ with minimal conductivity in $d=3$ RN-AdS, where $\mu$ is dimensionless chemical potential. 
  \label{fig3}}
\end{figure}

 \begin{itemize}
\item {\it Zero Charge Limit: }
Remember that the charge density in the boundary theory  is related to the 
chemical potential by 
\be\rho=\mu r_0^{d-2}/g^{2}\ .\ee When $\mu=0$ chargeless case, the $d=3$
conductivity is independent of the cutoff position. 
The flow solution for
$d=3$  is trivial while there are nontrivial flow solutions for $d\neq 3$~\cite{Sin:2011yh}. When $d\neq 3$
there is  a fixed point at boundary but not at horizon. 
This is because  near the boundary metric is aysmptotically AdS and scale invariance is there.  This scale invariance is lost when two fluctuating modes mix in the RN-AdS black hole case. 

\item {\it Extremal Limit: }
As shown in Figure \ref{fig1}, when the charged AdS black hole becomes extremal, one can find a fixed point from flow of
conductivity near the horizon due to the appearance of  AdS$_2$ near the horizon.  This fixed point will disappear in non-extremal case. One should notice that there is no fixed point for  the conductivity 
near the boundary both for extremal and non-extremal case. The evolution equation for $\sigma_A$ loses scale invariance near the infinite boundary in the presence of charge due to the mix of  $a_t,A_x$ modes.

\item {\it  Check against Boundary Result:}
In order to check the consistence with previous results calculated at infinite boundary $r=\infty$, we plot the boundary AC conductivity in Figure \ref{fig3},
which is precisely consistent with the results in~\cite{Hartnoll:2009sz}.
The $Q$ dependent term in the flow equation clearly give the origin for the   divergence of  the imaginary part of the conductivity in the zero frequency limit. 
 
\item  {\it Conductivity Minimum: }
It is interesting to observe that there is a window of parameters $\omega$ and $\mu$ where the real part of conductivity curve in radial direction has a minimum as  shown in Figure \ref{fig1}. The minimum of the 
conductivity pick up a certain scale $r_*$ and we show the charge dependence of that  scale in Figure $\ref{fig3}$. 
 \end{itemize}

\subsection{RG flow of `Conductivity'  of  momentum current  flow at zero momentum}
Since the metric perturbation  $a_\mu$ mimic the   Maxwell system, 
we can define the ``conductivity'' for the  mode $a_z=h^x_z$ by
\be
\sigma_h := {j^z\ov f_{zt}}\ .
\ee
At $k_z=0$, since $a_\mu$ modes are decoupled with $A_x$, we can obtain the decoupled flow of $\sigma_h$ as follow \be  {\p_r \sigma_h\ov -i\omega} +\sigma_h^2\ {g_{rr}g_{zz}\ov
G} - Gg^{tt}g^{zz} = 0\ . \ee
This
flow equation has the same form as the longitudinal conductivity flow in zero charge case although the metric components contain   $Q$ dependence. As a consistent check, this flow equation is exactly the same as shear viscosity flow in (\ref{etaflow}), because when $k_z=0$, there is no polarization direction and $\sigma_h$ is nothing but shear viscosity by definition. We will explain more about the physical mean of $\sigma_h$ in next subsection.

\subsection{ RG flow of `Conductivity'  of  momentum current in diffusion region} 
Since shear parts of metric perturbations behaves as longitudinal Maxwell field
with a $r$ dependent effective coupling, we shall work out the ``conductivity'' flow in diffusion region. In order to
handle the equations of motion: (\ref{heom0})$\sim$(\ref{heom3}),  
we write them in momentum
space and take the diffusion scaling following~\cite{Andy.RG} \bea
\p_t&\sim&\ep^2\ , \quad \p_z\sim\ep\\
f_{zt}&\sim& \ep^3 \left( {f_{zt}}^{(0)} + \ep {f_{zt}}^{(1)} + \dots\right)\ . \eea First consider the in-falling
boundary condition, which requires $j^z$ linearly related to $f_{zt}$ near the horizon. This condition requires us to
scale $j^z$ as same power of $f_{zt}$: \be j^z\sim \ep^3 \left( {j^z}^{(0)} + \ep {j^z}^{(1)} + \dots\right)\ . \ee
Through the charge conservation equation we obtain \be \tilde{j^t}\sim \ep^2 \left( \tilde{j^t}^{(0)} + \ep
\tilde{j^t}^{(1)} + \dots\right)\ , \ee and we can also obtain \be {j^t}\sim \ep^2 \left( {j^t}^{(0)} + \ep {j^t}^{(1)}
+ \dots\right)\ , \ee since there is no different scaling between $j^t$ and $\tilde{j^t}$. Now we want to take
$\ep\rightarrow 0$ limit, and only the lowest orders of fields are left. From ($\ref{heom1}$) and ($\ref{heom2}$) we obtain
\be \p_r\tilde{j^t}^{(0)} = 0\ ,\quad \p_r{j^z}^{(0)} = 0\ . \ee Requiring charge conservation, we have \be {j^z}^{(0)}
= \tilde{j^t}^{(0)}{k\ov \omega}\ . \ee From ($\ref{heom3}$), we have \be \p_rf^{(0)}_{zt} = {g_{rr}g_{tt}\ov
G}\p_z{j^t}^{(0)}\ . \ee It means we have the solution for $f^{(0)}_{zt}$ \be\label{fflow} f^{(0)}_{zt}(r) =
f^{(0)}_{zt}(r_0) + 2\kappa^2\int_{r_0}^r dr{g^2\p_z{j^t}^{(0)}(r)\ov \sqrt{-g}g^{rr}g^{tt}g_{xx}}\ . \ee
Following  the
usual definitions in sliding membrane for chargeless case,
we define the conductivity by current and electric field as \be \sigma_h (r_c) :=
{{j^z}^{(0)}(r_c)\ov f^{(0)}_{zt}(r_c)} = {k\ov\omega}{\tilde{j^t}^{(0)}(r_c)\ov  f^{(0)}_{zt}(r_c)}\ . \ee Remember
\be \tilde{j^t}^{(0)} =  {j^t}^{(0)} - \frac1{g^2}\sqrt{-g}A^{(0)}_x\bar A'_t \ee is a constant.  By taking the scaling limit we have the
simplified equation of motion for $A^{(0)}_x$: \be \p_r(\frac1{g^2}\sqrt{-g}g^{rr}g^{xx}\p_r A^{(0)}_x) = 
g^2_{\text{eff}} \bar A'_t{j^t}^{(0)}(r)\ . \ee The solution for $
A^{(0)}_x$ can be solved explicitly as shown in~\cite{Andy.RG}.

Since ${j^z}^{(0)}$ is
a constant in radial direction, after having the solution for $A^{(0)}_x$, (\ref{fflow}) can be rewritten as \be\label{Dsigmaflow} {1\ov \sigma_h(r_c)} = {1\ov\sigma_h(r_0)}
-{k_c^2\ov i\omega_c}{\bar D_h(r_c)/T_c\ov \sigma_h(r_0)}\ , \ee where the horizon value is
\be
\sigma_h(r_0)={1\ov 16\pi G_N}\left({r_0\ov
L}\right)^{d-1}\  .
\ee This is exactly the ``conductivity'' flow for shear part of metric fluctuations. This is our main result in this subsection. After taking the scaling limit following~\cite{Andy.RG}, we have the analytical result for RG flow of  ``conductivity''  of momentum current.  

As a byproduct, we obtain the $\bar D_h$
as \be\label{QD}\bar D_h(r_c) = {1\ov 4\pi} {(d-2)(1-\alpha Q^2) +2 \ov (1+\alpha Q^2)d-2(d-1)\alpha
Q^2(r_0/r_c)^{d-2}}\ ,\ee which is charge dependent diffusion constant.  Notice that we have
used the solution for $ A_x$ to obtain the $\bar D_h$. This result was first obtained in~\cite{Andy.RG}, where the diffusion constant is derived from Fick's law, while here we find that this diffusion constant is included in conductivity flow for momentum current without any boundary condition for $f^{(0)}_{zt}$ in ($\ref{fflow}$).

Now we will explain the physical mean of the ``conductivity'' $\sigma_h$. Notice that $a_z\equiv \delta g^x_z$
corresponds to boundary operator $T_x^z$. At zero frequency retarded Green function of $T_x^z$ is exactly the same as
$T_x^y$ since there is no special polarization direction. Thus the transport coefficient given by $a_z\equiv\delta
g^x_z$ will correspond to longitudinal momentum dependent viscosity. This can be confirmed at the horizon, since
$\sigma_h$ at horizon is nothing but~\cite{Hong.Membrane} \be \sigma_h(r_0) = {s\ov 4\pi}\ ,\ee which equals to shear
viscosity. This is consistent with $\eta/s={1\ov 4\pi}$. One should note that transport coefficients at the horizon are all frequency
independent.

\subsection{Mixed RG Flow Equations}
We shall discuss the mixed flows coming from coupled equations of motion without any scaling limit.
Consider again the coupled equations of motion: (\ref{heom0}) - (\ref{heom3}) and (\ref{Aeom}).
 By defining \be\label{sigmadef} \sigma_h = {j^z\ov f_{zt}}\ ,\quad \sigma_A = {J^x \ov i\omega A_x}\ , \quad \alpha := {j^z\ov A_x}\ 
\ee we can derive the following flow equation for them as follows
\\[2mm]
\be\label{flowsigmahF}
{\p_r \sigma_h\ov -i\omega} + \sigma_h^2\left[{g_{rr}g_{zz}\ov G} -{g_{rr}g_{tt}\ov G}\left(-{k^2\ov \omega^2} +
\frac1{g^2}\sqrt{-g}g^{tt}g^{rr}\bar A'_t {k \ov \omega\alpha}\right) \right] - Gg^{tt}g^{zz} = 0\ , \ee 
 \be\label{flowsigmaAF}
{\p_r \sigma_A\ov -i\omega} + {g^2\sigma_A^2\ov \sqrt{-g}g^{rr}g^{xx}} +{g^2_{\text{eff}}\bar A'_t\ov \omega^2}\left({k\alpha\ov \omega} + \frac1{g^2}\sqrt{-g}\bar A'_t\right) 
+ \frac1{g^2}\sqrt{-g}g^{xx}\left(-g^{tt} + g^{zz}{k^2\ov \omega^2}\right) = 0\ , \ee
\be\label{flowalpha} {\p_r\alpha\ov -i\omega} = 
\alpha\left({Gg^{tt}g^{zz}\ov \sigma_h} - {g^2\sigma_A\ov \sqrt{-g}g^{xx}g^{rr}}\right)\ . \ee
One can easily observe that when $Q=0$, all mixing effects disappear and  ($\ref{flowsigmahF}$) and ($\ref{flowsigmaAF}$) will reduce to
longitudinal and transverse form of conductivity flow for  chargeless case~\cite{Hong.Membrane} respectively. 
Eqs. (\ref{flowsigmahF}), (\ref{flowsigmaAF}) and (\ref{flowalpha}) give the mixed RG flow.

\section{ hydrodynamics at the finite holographic screen}\label{hydro}
In this section, we study the RG flow of
transport coefficients by calculating
the Green functions at finite screen  $r=r_c$.
We consider the shear mode
in the 5 dimensional RN-AdS background.
This hydrodynamics problem is considered in \cite{Ge:2008ak} for $r_c \to \infty$ case.
We use the following gauge 
\be
 a_r = 0\ , \quad A_r = 0\  
\ee
and introduce following coordinate and notations:
\be
 u = \frac{r_0^2}{r^2}\ ,\quad
 a = \alpha Q^2\ ,\quad
 b = \frac{L^2}{2r_0}\ .
\ee
\noindent
In terms of the rescaled vector $ B(u) =  {A_x(u)}/{\mu}$, 
the equations of motion are given by
\begin{subequations}
\begin{align}
0 &= {a_t}''-\frac{1}{u}{a_t}' -\frac{b^2}{uf}\Big(\omega k a_z+k^2 a_t\Big) -3auB', \label{eq_motion_v_001}
\\
0
&=
kf{a_z}'+\omega{a_t}'-3a\omega uB,
\label{eq_motion_v_002}
\\
0
&=
{a_z}''+\frac{(u^{-1}f)'}{u^{-1}f}{a_z}'
+\frac{b^2}{uf^2}\Bigl(\omega^2 a_z+\omega k a_t\Bigr),
\label{eq_motion_v_003}
\\
0
&=
B''+\frac{f'}{f}B'+\frac{b^2}{uf^2}\Bigl(\omega^2-k^2f\Bigr)B
-\frac{1}{f}{a_t}'.
\label{eq_motion_v_004}
\end{align}
\end{subequations}
Notice that $B$ and $a_t$ couples which is the source of the complication.
To handle the problem, we introduce the master fields
\cite{Kodama:2003kk},
\begin{equation}
\Phi_\pm
=
\frac{1}{u}{a_t}'-3aB+\frac{C_\pm}{u}B, \label{Phipm}
\end{equation}
with  $C_\pm$ given by
$$
C_\pm
=(1+a)\pm\sqrt{(1+a)^2+3ab^2k^2} .
$$
The decoupled differential equations  satisfied by the master fields are  
\begin{equation}
0=
{\Phi_\pm}''
+\frac{(u^2f)'}{u^2f}{\Phi_\pm}'
+\frac{b^2}{uf^2}\Big(\omega^2-k^2f\Big)\Phi_\pm
-\frac{C_\pm}{f}\Phi_\pm .
\label{master_eq_01}
\end{equation}
We expand the master fields  in the   hydrodynamic regime in following way:%
\footnote{
Here, the definition of $F_{+1}(u)$ is related to
the $\widetilde F_1(u)$ in \cite{Ge:2008ak} as
$$
 b F_{+1}(u) = \widetilde F_1(u) -\frac{i}{4\pi T}\log(1-u) ,
$$
and similarly for $F_{-1}(u)$, etc.
}
\begin{align}
 \Phi_+
 &=
  g(u) {\widetilde C}
 (1+b \omega F_{+1}(u) + b^2 k^2 G_{+2}(u) + b^2 \omega^2 F_{+2}(u) + \cdots)
\\
 \Phi_-
 &=
 C (1+ b \omega F_{-1}(u) + b^2 k^2 G_{-2}(u) b^2 \omega^2 F_{-2} (u) + \cdots) ,
\end{align}
where the factor singular near the boundary is explicitly taken out for $\Phi_+$ and it is 
given by 
\begin{equation}
 g(u) = \frac{1}{u} -\frac{3a}{2(1+a)} .
\end{equation}
In this regime, the equations of motion are solved
by imposing the ingoing boundary condition at the horizon
in \cite{Ge:2008ak}.
Here, we  collect a few terms which will be relevant later
\begin{align}
 F_{+1}'(u)
 &= i\frac{(2-a)^2}{4(1+a)^2}\frac{1}{u^2f(u)g^2(u)} ,
 \label{F'+1}
 \\
 F_{-1}'(u)
 &= \frac{i}{u^2 f(u)} , \label{F'-1}\\
 G_{-2}'(u) &= -\frac{1}{2(1+a)u^2} . \label{F'-2}
\end{align}
The constants $C$ and $\widetilde C$ are integration constants and fixed by imposing the boundary conditions. 
It is convenient to define a gauge invariant field $Z $ by 
\begin{align}
 k a_t(u) + \omega a_z(u ) &= Z  .
 \end{align}
 The gauge field $B(u)$ can be expressed in terms of master fields as 
\begin{equation}
 B
 =
\frac{1}{C_+-C_-}u\Bigl(\Phi_+-\Phi_-\Bigr).
\label{B}
\end{equation}
Using \eqref{eq_motion_v_001} and \eqref{Phipm}, 
we obtain 
\begin{equation}
 u^2\Phi_\pm'
-C_\pm uB'
=\frac{b^2}{f} kZ
-C_\pm B. \label{BasicRelation}
\end{equation}
With \eqref{B}, l.h.s. of \eqref{BasicRelation} can be expressed
in terms of master fields $\Phi_\pm$. Requiring (\ref{BasicRelation}) at $u=u_c$ we can determine $C$ and $\tilde C$ in terms of the boundary values of 
$Z$ and $B$  at $u=u_c$ :
\begin{align}
 C &=
 \frac{\alpha(u_c)b^2 k Z_c - \beta(u_c)f(u_c) B_c}
 {u_c g(u_c)(-i\omega + D(u_c) k^2)b} , \\
 \widetilde C &=
 \frac{\widetilde\alpha(u_c)b^2 k Z_c - \widetilde\beta(u_c)f(u_c)B_c}
 {u_c g(u_c)(-i\omega + D(u_c) k^2)b} ,
\end{align}
where denominators are expressions up to $\mathcal O(\omega^2)$
and $\mathcal O(k^3)$, and
$D(u_c)$ is given by
\begin{equation}
 D(u_c) = \frac{b f(u_c)}{2(1+a) u_c g(u_c)} = \frac{b f(u_c)}{2(1+a)-3au_c}\ .
\end{equation}
The coefficients $\alpha$, $\beta$,
$\widetilde\alpha$ and $\widetilde\beta$ are
given in terms of the solutions at $u=u_c$ as
\begin{align}
 \alpha(u_c)
 &=
 -u_c g(u_c) \left(1 + b \omega F_{+1}(u_c) + b^2 k^2 G_{+2}(u_c)
 \right) +\cdots \\
 \beta(u_c)
 &=
 \frac{3ab^2 k^2}{2(1+a)} + \cdots\\
 \widetilde\alpha(u_c)
 &=
 - i b \omega + \frac{b^2 k^2}{2(1+a)}
 + \cdots
 \\
\widetilde\beta(u_c)
 &=
 2(1+a) i b \omega - b^2 k^2 \cdots .
\end{align}

Now we calculate the Green functions at $u=u_c$.
We start from the Einstein-Hilbert action with
Gibbons-Hawking terms and counter terms \cite{Balasubramanian:1999re}
for the gravity part~\footnote{We use this action in order to keep consistence with previous results when $u_c\rightarrow 0$.}
\be
S_\text{gravity} = S_{\text{EH}} + S_{\text{G.H}} + S_{c.t}\ ,
\ee
where the Gibbons-Hawking terms and counter terms can be expressed 
in terms of (trace of) the extrinsic curvature $K$ and 
induced metric $\gamma_{\mu\nu}$ as 
\bea
S_\text{EH} &=& -{1\ov 16\pi G}\int d^{d+1}x \sqrt{-g} R\ ,\nn
S_\text{G.H} &=& -{1\ov 8\pi G}\int d^dx \sqrt{-\gamma} K\ , \nn
S_{c.t}& =& {1\ov 8\pi G}\int d^d x\sqrt{-\gamma}{3\ov L}\ .
\eea
Both Gibbons-Hawking term and counter terms are defined as the hypersurface at $u=u_c$. After perturbing the action and integrating out the classic solution of perturbations, finally we obtain
the boundary action for the shear modes, which are given by
\begin{align}
 S^{o.s}_\text{gravity}
  &=
 \frac{L^3}{32\kappa^2b^4}\int d^4x\biggl[
  \frac{1}{u} a_t a'_t
  - \frac{3}{u^2}\left(1-\frac{1}{\sqrt{f(u)}}\right) a_t a_t
  \notag\\&\quad
  - \frac{f(u)}{u} a_z a'_z
  + \frac{1}{u^2}\left(3f(u)-3\sqrt{f(u)}-u f'(u)\right) a_z a_z
 \biggr]\ ,
\\
 S^{o.s}_\text{gauge}
 &=
 \frac{L^3}{32\kappa^2b^4}\int d^4x\left[
 -3a f(u) B B' + 3 a B a_t \right] .
\end{align}
Once we have $C\ , \tilde C$ in terms of boundary values, by the definitions of master fields we can express
first derivatives of $a_t$, $a_z$ and $B$
in terms of boundary values at $u=u_c$ as follows:
\begin{align}
 a_t'(u_c)
 &=
 \frac{\alpha_t(u_c)b^2 k Z_c - \beta_t(u_c)f(u_c) B_c}
 {u_c g(u_c)(-i\omega + D(u_c) k^2)b} \\
 a_z'(u_c)
 &=
 \frac{\alpha_z(u_c)b^2 k Z_c - \beta_z(u_c)f(u_c) B_c}
 {u_c g(u_c)(-i\omega + D(u_c) k^2)b} \\
 B'(u_c)
 &=
 \frac{\alpha_B(u_c)b^2 k Z_c - \beta_B(u_c)f(u_c) B_c}
 {u_c g(u_c)(-i\omega + D(u_c) k^2)b} \ ,
\end{align}
where the coefficients are explicitly given by
\begin{align}
 \alpha_t(u_c)
 &=
 - u_c^2 g(u_c)
 \left[1+b \omega (F_{+1}(u_c)+F_{-1}(u_c)))
 +b^2k^2(G_{+2}(u_c)+G_{-2}(u_c))\right]
 + \cdots ,\\
 \beta_B(u_c)
 &=
 -\frac{3a ib\omega}{2(1+a)f(u_c)} + \cdots\ ,
\end{align}
where $\dots$ denotes higher frequency and high momentum terms. Notice that we only write down two coefficients in above equations for later use and leave other coefficients in Appendix B.  With the help of above results, the Green functions at slice $u=u_c$ can be read off from on shell action
\begin{align}
 G_{xt\,xt}
 &=
 \frac{L^3}{16\kappa^2b^4}\biggl\{
 \frac{\alpha_t(u_c) b^2 k^2}
 {u_c^2 g(u_c)\left(-i\omega+ D(u_c)k^2\right)b}
 -\frac{3}{u_c^2}\left(1-\frac{1}{\sqrt{f(u_c)}}\right)
 \biggr\} ,
\\
 G_{xz\,xz}
 &=
 \frac{L^3}{16\kappa^2b^4}\left\{
 \frac{\alpha_t(u_c) b^2 \omega^2}
 {u_c^2 g(u_c)\left(-i\omega+ D(u_c)k^2\right)b}
 + \frac{1}{u_c^2}\left(3f(u_c)-3\sqrt{f(u_c)}-u_c f'(u_c)\right)
 \right\} ,
\\
 G_{xt\,xz}
 &=
 \frac{L^3}{16\kappa^2b^4}
 \frac{2\alpha_t(u_c) b^2 k \omega}
 {u_c^2 g(u_c)\left(-i\omega+ D(u_c)k^2\right)b} ,
\\
 G_{xt\,x}
 &=
 \frac{L^3}{32\kappa^2b^4\mu}\left\{
 \frac{-\beta_t(u_c)f(u_c)- 3 a u_c f(u_c) \alpha_B(u_c)b^2 k^2}
 {u_c^2 g(u_c)\left(-i\omega+ D(u_c)k^2\right)b}
 + 3 a
 \right\} ,
\\
 G_{xz\,x}
 &=
 \frac{L^3}{32\kappa^2b^4\mu}
 \frac{\beta_z(u_c) f^2(u_c) - 3 a u_c f(u_c) \alpha_B b^2 k \omega}
 {u_c^2 g(u_c)\left(-i\omega+ D(u_c)k^2\right)b} ,
\\
 G_{x\,x}
 &=
 \frac{L^3}{16\kappa^2b^4\mu^2}
 \frac{3 a f^2(u_c)\beta_B(u_c)}
 {u_c g(u_c)\left(-i\omega+ D(u_c)k^2\right)b}\ .
\end{align}

\subsection{Cut off dependence of diffusion constant}
One can easily observe a universal diffusion constant depending on the cutoff position from all the Green functions:
\begin{equation}
 D(u_c) = \frac{b f(u_c)}{2(1+a) u_c g(u_c)} = \frac{b f(u_c)}{2(1+a)-3au_c}\ .
\end{equation}
Change to orthonormal frame, one can obtain
\be \hat D(u_c) = D(u_c){g_{zz}\ov \sqrt{g_{tt}}}\ . \ee
Together with a normalization with local temperature, one can obtain the dimensionless diffusion constant
\bea
\bar D_c &=& \hat D(u_c) T_c = D(u_c)T_H \left({g_{zz}\ov g_{tt}}\right) \nn
&=&  {1\ov 4\pi} {2-a \ov 2(1+a)-3au}\ .
\eea
This is nothing but $\bar D_h$ in (\ref{QD}). The above result was first obtained in~\cite{Andy.RG} by taking a certain scaling for the equations of motion. Here we show that this diffusion pole appears in all the cut-off dependent Green functions.

\subsection{Shear viscosity}
The shear viscosity is calculated  by using Kubo formula
\begin{equation}
 \eta(r_c) = -\lim_{\omega\to 0}\frac{\mathrm{Im}G_{xy\,xy}(\omega,k=0, r_c)}{\omega} .
\end{equation}
For $k=0$, $z$-direction can be treated equivalently to $y$-direction, since there is no polarization direction, and we have
\begin{equation}
 G_{xy\,xy}(\omega,k=0, r_c) = G_{xz\,xz}(\omega, k=0, r_c)
  = -i\frac{L^3}{16\kappa^2 b^3}\omega + \mathcal O(\omega^2) .
\end{equation}
Then the shear viscosity becomes
\begin{equation}
 \eta(r_c) = {1\ov 16\pi G_N}\left({r_0\ov L}\right)^3 ,
\end{equation} which is constant independent of cut-off. This is consistent with the result from flow equation (\ref{floweta}).
\subsection{Cut off dependence of DC conductivity}
The conductivity can be calculated form
the $\mathcal O(\omega)$ term of the
Green function $G_{x\,x}$.
Generally it contains the $\mathcal O(\omega^2)$ terms of
the master fields and have a complicated expression.
Taking into account the $\mathcal O(\omega^2)$ terms,
it becomes
\begin{equation}
  G_{x\,x}
 =
 \frac{L^3}{16\kappa^2b^4\mu^2}
 \frac{3 a f^2(u_c)\beta_B(u_c)}
 {u_c g(u_c)\left(-i\omega+ D(u_c)k^2
         -u_c^2f(u_c)(F_{-2}'+F_{+1}'F_{-1}')\right)b} ,
\end{equation}
where $\beta_B(u_c)$ also have the $\mathcal O(\omega^2)$ corrections as
\begin{equation}
 \beta_B(u_c) = -\frac{3a ib\omega}{2(1+a)f(u_c)}
 + b^2 \omega^2 u^2 \left[\left(u g(u)\right)' F_{-2}'(u)
 + \left(u g(u) F_{+1}\right)' F_{-1}'(u) \right]
\end{equation}
For $k\to 0$, the expression of $G_{x\,x}$ is simplified and given by
\begin{align}
 G_{x\,x}
 &=
 \frac{3 a f(u_c) L^3}{16\kappa^2b^4\mu^2}
 \left(\frac{\left(u_c g(u_c)\right)'}{u_c g(u_c)}
  + b\omega F_{+1}'(u_c)\right) + \mathcal O(\omega^2)
 \notag \\
 &=
 \frac{3 a f(u_c) L^3}{16\kappa^2b^4\mu^2}
 \left(\frac{\left(u_c g(u_c)\right)'}{u_c g(u_c)}
  + ib\omega\frac{(2-a)^2}{4(1+a)^2}\frac{1}{u_c^2f(u_c)g^2(u_c)} \right)
 + \mathcal O(\omega^2) \ .
\end{align}
The real part electric conductivity is explicitly given by Kubo formula
\be\label{flowDC}
\sigma^c_{DC} = {1\ov g^2}{r_0\ov L} {(2-a)^2\ov 4(1+a)^2}{1\ov u_c^2 g(u_c)^2}\ ,
\ee where $g$ is the gauge coupling.
At the horizon, $\sigma_{DC}$ is
\be
\sigma_{H} =  {1\ov g^2}{r_0\ov L} = {1\ov g^2}\frac{\pi L}{2}(T+\sqrt{T^2+\mu^2/3\pi^2}),
\ee
which is consistent  with  eq.(58) in~\cite{Hong.Membrane} in the limit
where charge or chemical potential goes to 0.
This horizon value is related to the
membrane conductivity ${\hat \sigma_H}=J^i_{mb}/{\hat E^i}$
 \cite{Hong.Membrane}  by
\be
{\hat \sigma_H}=\frac{\sqrt{g_{tt}g_{rr}}}{\sqrt{-g}} g_{ii}\cdot \sigma_H =  {1\over g^2}.\ee   One can also check that as one goes to
the boundary ($u_c\rightarrow 0$), DC conductivity is reduced to
\be
\sigma_{DC} = {1\ov g^2}{r_0\ov L} {(2-a)^2\ov 4(1+a)^2}\ ,
\ee
which  precisely agrees with previous result in~\cite{Ge:2008ak}.

\begin{figure}[hbtp]
\includegraphics*[bb=0 0 250 160,width=0.45\columnwidth]{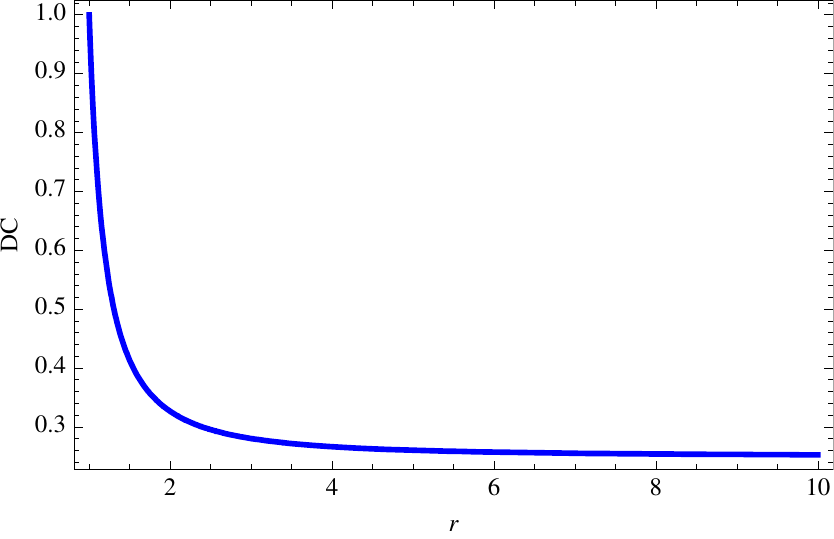}
~~~~~~~~~
\includegraphics*[bb=0 0 250 160,width=0.45\columnwidth]{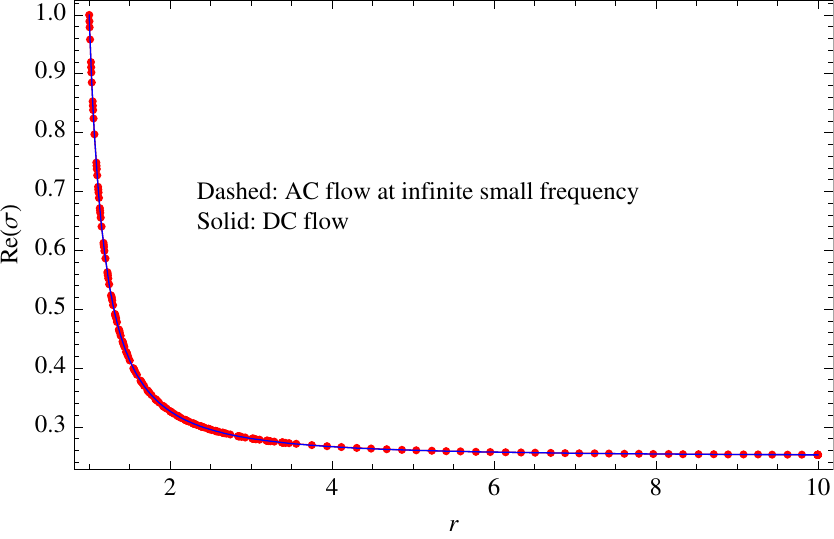}
\caption{ Checking consistency between flows of DC conductivities: Left: Plot of equation (\ref{flowDC}) from hydrodynamics. Right: comparing the hydrodynamic result and  flow equation result. We can see they are explicitly same with each other.
  \label{Check}}
\end{figure}

Finally we can check the consistency of flow equation of AC conductivity
by comparing its numerical value in the limit of zero frequency
with that of DC conductivity calculated here.
The result is plotted in figure \ref{Check}.
\\[1mm]

\section{Conclusion and Discussion}
In this paper, we present a method to work out the RG flows for transport coefficients for quark gluon plasma at finite chemical potential with 
charged AdS black hole dual.  Due to the mixing effect between Maxwell and metric perturbations, we need to solve the coupled equations of motion, which is usually difficult. We organize the system as two coupled Maxwell systems and define two conductivities for each of them. With a parameter characterizing the mixing effect, we write down the mixed flow equations.  These mixed RG flow equations will be simplified in certain limits. These flow equations will characterize how the transport coefficients will change as energy scale changes.

 We explicitly give the flow equations for conductivity and shear viscosity. In order to check these results analytically we use hydrodynamic method to calculate the Green function at finite cut-off slice $r=r_c$. We impose equations of motion at $r=r_c$, which is guaranteed by RG invariance of bulk action at classical level~\cite{Sin:2011yh}. Then the Green function can be read off from the on-shell action at $r=r_c$. We extended the usual counter term to arbitrary slice in order to have consistent result when $r_c\rightarrow \infty$. By using Kubo formula we obtain the analytical results of RG flow formulas for transport coefficients and we found complete agreements with that from flow equations.

\section*{Acknowledgements}
This work was supported by the National Research Foundation of Korea(NRF) grant funded by the
Korea government(MEST) through the Center for Quantum Spacetime(CQUeST) of Sogang University with grant number
2005-0049409. SJS was also supported by Mid-career Researcher Program through NRF grant (No. 2010-0008456 ). 
YM is supported by JSPS Research Fellowship for Young Scientists and 
in part by Grant-in-Aid for JSPS Fellows (No.23-2195).

\appendix
\section{Derivation of bulk action}
 Starting from Einstein-Hilbert action, we calculate the bulk action for vector perturbations $a_z, a_t, a_r, A_x$ in the 5D RN-AdS background. 
 \be
 S = \int d^5x\biggr [A_1 a^\prime_t a^\prime_t +   
  A_2 a^\prime_z a^\prime_z  
 + D A^\prime_xA^\prime_x + E a_t A^\prime_x\ +\left(\half B_1 a_t a_t\right)' + \left(\half B_2 a_z a_z\right)' +\cdots 
 \biggr].
 \ee
Here all the prime denotes the $r$ derivative and $\cdots$ represent the terms with $\partial_t,\partial_z$. 
The coefficients are given by
\bea
A_1&=& {1\ov 2\kappa^2}{r^5\ov 2L^5}, \quad A_2= {1\ov 2\kappa^2}{-f r^5\ov 2L^5}\\
B_1&=& {1\ov 2\kappa^2}{r^4\ov L^5}\left(-2+{rf^\prime\ov f}\right), \quad
B_2= {1\ov 2\kappa^2}{2r^4f\ov L^5}\\ 
D&=& {1\ov g^2}{-r^3f\ov 2L^3}, \quad E= {1\ov g^2}{r^3{\bar A}^\prime_t\ov L^3}\ .
\eea
Notice that the total derivative terms can be deleted for the purpose of the 
equations of motion.
One can evaluate equations of motion from the action for $a_t, a_z, A_x$:
\bea
 A_1 a^{\prime\prime}_t + A^\prime_1 a^\prime_t + E A^\prime_x&=&0\\
A_2 a^{\prime\prime}_z + A^\prime_2 a^\prime_z &=&0\\
(DA^\prime_x + E a_t)^\prime &=& 0.
\eea  However we should calculate the final on-shell action by adding proper local counter terms, as we did in section \ref{hydro}. 
We conclude that the mixed bulk action can be written equivalently as two Maxwell actions plus the only mixing term $ E a_t A^\prime_x$.

\section{Coefficients for $a_t'$, $a'_z$ and $B'$}
The coefficients for $a_t'$, $a'_z$ and $B'$ are given by
\begin{align}
 \alpha_t(u_c)
 &=
 - u_c^2 g(u_c)
 \left[1+b \omega (F_{+1}(u_c)+F_{-1}(u_c)))
 +b^2k^2(G_{+2}(u_c)+G_{-2}(u_c))\right]
 + \cdots ,
 \\
 \alpha_z(u_c)
 &=
 -\frac{\omega}{k f(u_c)}\alpha_t(u_c)
 \\
 \alpha_B(u_c)
 &=
 -\frac{u^2 g'(u_c)}{2(1+a)}
 + \frac{b u_c^2 w \left(g(u_c)
   \left(F_{-1}'(u_c)-F_{+1}'(u_c)\right)-(F_{-1}(u_c)+F_{+1}(u_c))
   g'(u_c)\right)}{2 (a+1)}
 \notag\\&\quad
 +\frac{b^2 k^2 u_c^2 \left(g(u_c)
   \left(G_{-2}'(u_c)-G_{+2}'(u_c)\right)-(G_{-2}(u_c)+G_{+2}(u_c))
   g'(u_c)\right)}{2 (a+1)}
 \notag\\&\quad
 + \frac{3au_c^2 g'(u_c)b^2 k^2}{4(1+a)^3} + \cdots , \\ 
 \beta_t(u_c)
 &=
 ib\omega\frac{3au_c^2 g(u_c)}{f(u_c)} + \cdots ,
 \\
 \beta_z(u_c)
 &=  -\frac{3 a u_c b^2 k \omega}{2(1+a)} + \cdots , \\
 \beta_B(u_c)
 &=
 -\frac{3a ib\omega}{2(1+a)f(u_c)} + \cdots.
\end{align}

\end{document}